# Excel 2013 Spreadsheet Inquire

Patrick O'Beirne
Systems Modelling Ltd
pob@sysmod.com

## ABSTRACT

*Excel 2013 (version 15) includes an add-in 'Inquire'[1] for auditing spreadsheets. We describe the evolution of such tools in the third-party marketplace and assess the usefulness of Microsoft's own add-in in this context. We compare in detail the features of Inquire with similar products and make suggestions for how it could be enhanced. We offer a free helper add-in that in our opinion corrects one major shortcoming of Inquire.*

## INTRODUCTION

From the earliest days of spreadsheets, auditors and reviewers have been seeking ways of gaining an understanding of their underlying structure of formulas. Once Excel arrived and large worksheets became common, it rapidly became obvious that simple formula listings were too tedious to work with, and spreadsheet auditing tools were developed that provided two indispensable services: a visual overview of the structure of a spreadsheet, and a detailed listing of features ('bad smells') which from the experience of the auditors may indicate an actual error or a weakness that could easily materialise as a defect in use.

One of the first of these, and still in use, is the Spreadsheet Detective[2] which created maps using single-letter codes such as F for Formula, and symbols such as > to indicate a formula was copied to its right. With a small column width and a black-and-white printer, one could quickly see how a spreadsheet had been structured. Later, colouring was added tools and auditing software like SpACE[3] added many colourful schemes and detailed lists of suspect features. The authors of SpACE also published one of the first frameworks for spreadsheet audit, as presented by Ray Butler in a pioneering EuSpRIG 2000 paper " Risk Assessment For Spreadsheet Developments: Choosing Which Models to Audit"[4]. For a later view of how financial service companies do it, see "A Typical Model Audit Approach: Spreadsheet Audit Methodologies in the City of London" by Grenville Croll[5].

Many more such products have been created in what was largely a cottage industry. The author has his own product, XLTest[6]. Some added features to assist with building and debugging the formulas in certain types of spreadsheets, eg financial modelling, such as Operis OAK[7]. Some, with the impetus of the FDA's 21 CFR Part 11, focused on locking down spreadsheets under control, such as ABB's DACs[8]. More recently, enterprise spreadsheet management systems such as Liquidity[9], ClusterSeven[10] and Cimcon[11] have provided a visualisation of the use and development over time of systems of spreadsheets, and have added monitoring and control features. For example, see the white papers by Clusterseven or the EuSpRIG 2007 paper "Managing Critical Spreadsheets in a Compliant Environment by Soheil Saadat[12]. A list of these tools is maintained at http://www.sysmod.com/sslinks.htm

These tools are like standard static software analysis. While they considerably increase the efficiency of the auditor in getting through large amounts of routine analysis, they are less useful as a predictor of the cost of an audit. This is discussed by David Colver in his EuSpRIG 2011 paper "Drivers of the Cost of Spreadsheet Audit"[13]. Neither will they detect errors in logic. In their EuSpRIG 2010 paper "The Detection of Human Spreadsheet Errors by Humans versus Inspection (Auditing) Software", Salvatore Aurigemma and Raymond R. Panko reported "[tools] were almost useless for correctly flagging natural (human) errors in this study."

These tools are excellent in raising questions; the remedial actions are however a matter of skill and judgment, as discussed by Louise Pryor in her Eusprig 2008 paper "Correctness is not enough"[14].

## THE INQUIRE ADD-IN

Excel version 15 has an Add-in that came from Microsoft's purchase of the Prodiance SpreadsheetIQ product, which adds an "INQUIRE" tab to the Ribbon.

To activate the Add-in, click File > Options > Add-Ins, in the Manage list, select COM Add-ins, click the Go button, and check the Inquire Add-in.

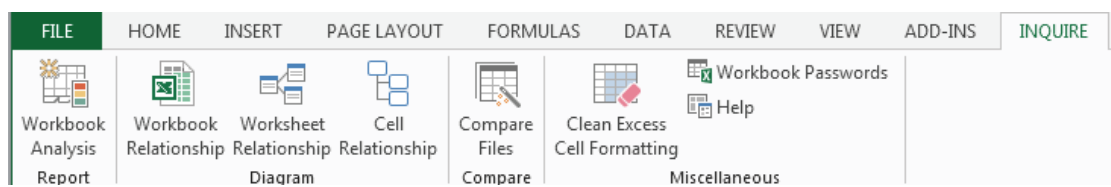

## OVERVIEW OF THE COMMANDS

The Workbook Analysis Report creates an interactive report showing detailed information about the workbook and its structure, formulas, cells, ranges, and warnings.

The Workbook Relationship Diagram creates an interactive, graphical map of workbook dependencies created by connections (links) between files. The types of links in the diagram can include other workbooks, Access databases, text files, HTML pages, SQL Server databases, and other data sources.

The Worksheet Relationship Diagram creates an interactive, graphical map of connections (links) between worksheets both in the same workbook and in other workbooks.

The Cell Relationship Diagram creates a detailed, interactive diagram of all links from a selected cell to cells in other worksheets or even other workbooks. These relationships with other cells can exist in formulas, or references to named ranges. The diagram can cross worksheets and workbooks.

The Compare Files command lets you see the differences, cell by cell, between two open workbooks. Results are colour coded by the kind of content, such as entered values, formulas, named ranges, and formats. A window shows VBA code changes line by line.

The Excel 2013 preview had an Interactive Diagnostics command that was withdrawn from the released version.

The Clean Excess Cell Formatting command removes excess formatting and can greatly reduce file size. This helps you avoid "spreadsheet bloat," which improves Excel's speed.

The Workbook Passwords command maintains a password list, which will be saved on your computer encrypted and only accessible by you. The password list is needed so that Inquire can open its saved copy of your workbook.

The Help command gives details and examples of these commands.

In the following section we describe these commands in detail, give some of their uses and show their limitations.

**WORKBOOK ANALYSIS REPORT**

If the workbook has unsaved changes, you will be asked "Only changes that have been saved will be included in this analysis. Would you like to save your changes now?" Click Yes to save the workbook.

The Summary sheet shows the basic properties: Workbook name, Creation Date, Modified Date, File Size (bytes), Title, and Author; and then the count of each of these detailed findings which are listed in separate subsequent sheets. It would be useful for the Summary sheet to be hyperlinked to the sheets listed.

| | |
|---|---|
| Linked Workbooks | DDE Links |
| Data Connections | Visible Sheets |
| Hidden Sheets | Very Hidden Sheets |
| Formulas | Array Formulas |
| Formulas With Errors | Formulas With Logical Values |
| Formulas With Numeric Values | Formulas With Date/Time Values |
| Formulas With Textual Values | Formulas With Numeric Constants |
| Formulas With Textual Constants | Formulas With Nested IF Statements |
| Formulas Without Cell References | Formulas Referencing Blank Cells |
| Formulas Referencing Hidden Cells | Formulas Referencing Text Cells |
| Formulas Referencing External Workbooks | Formulas Formatted As Text |
| Positive Formulas | Negative Formulas |
| Unique Formulas | Duplicate Formulas |
| Inconsistent Formulas | Cells With Dependents |
| Cells With Textual Constants | Cells With Numeric Constants |
| Cells With Comments | Cells With Validation Criteria |
| Cells With Conditional Formatting | Cells With Numerics Stored As Text |
| Invisible Cells | Used Input Cells |
| Unused Input Cells | Occupied Cells |
| Merged Cells | Blank Cells |
| Blank Referenced Cells | Unlocked Cells |
| Hidden Rows and Columns | Named Items |
| Named Items With Errors | Warnings |

**HOW TO USE THE SUMMARY REPORT**

You would scan this sheet and note items of interest, such as items that you would not expect to find, or to find in different numbers to what is reported. What you notice is going to be influenced by your previous experience and expectations of this spreadsheet, or similar spreadsheets from this user or business unit, and your knowledge of Excel in general. If the meaning of any item is not clear, you may need to refer to some training materials, or ask questions on online Excel forums, to obtain further information. It is beyond the scope of this paper to provide such explanation.

## WHAT IS MISSING FROM THE SUMMARY REPORT

The Summary sheet lists for the properties only the Title and Author.. Other tools show the name of the previous author, the last print date, and custom document properties which may indicate company-specific information such as contact people or tracking numbers.
It does not report whether the workbook was saved with Automatic or Manual calculation, nor the existence of Circular references or Iteration.
It does not report Styles – not even a count, which would indicate workbooks with excessive styles; nor unusual number formats that may hide data.
It does not report VBA – not even the presence of macros is shown.
Neither does it show the use of Data Lists, Pivot Tables, Consolidation Sources, Views or What-if features such as Scenarios or Goal Seek.

The detailed list sheets all begin with three rows of two merged cells at the top which means that if you copy and paste either the heading in A1 or the filename in A2 into a Word document, or into a worksheet as values, you get two copies.

## THE REMAINING SHEETS IN THE WORKBOOK ANALYSIS REPORT

The lists of links (Linked Workbooks, DDE Links, Data Connections) give only the bare name or link reference, not where they were found, and do not indicate whether the link source file actually exists, nor whether the linked content is up-to-date, nor for queries where they are used nor the CommandText.

There are separate sheets to list the names of Visible Sheets, Hidden Sheets, and Very Hidden Sheets, but it does not indicate the presence of Chart sheets, nor of Macro or Dialog sheets which may exist in very old Excel files. It does not gives the following information which other tools report such as excessive used range; the Print Area; Page Header and Footers; Objects such as embedded charts or graphics or Form or ActiveX buttons; whether a worksheet has the old Lotus evaluation rules set; whether the sheet windows are not displaying the Formula Bar, Scroll Bars, Row & Column headings; Zero values; and user-definable search strings such as "password".

The 'All Formulas' sheet as one per line in A1 style, with no concept of how the same formula in relative terms has been copied over areas. This is to me the single biggest failing. The vast majority of spreadsheets contain rows, columns, or blocks of repeated formulas, also known as clones or copies of a unique formula. One of the most important tests of the structural soundness of a spreadsheet is to verify the integrity of such areas, that they are neither too small nor too large relative to similar areas, and that no inconsistent entries exist inside them. This has been a feature of nearly every other spreadsheet formula audit tool. For a detailed discussion of the mechanics see the EuSpRIG 2008 paper by Markus Clermont "A Spreadsheet Auditing Tool Evaluated in an Industrial Context"[15].

I find it hard to believe that such a list, with an implication that each formula is different and has to be individually checked off in a list, would have been accepted as useful in a commercial product. Their list of "unique formulas" is therefore exactly the same as the complete formula list, except for any formulas which are duplicated in A1 style. Therefore we make available an IQHelper add-in to group similar formulas together to make checking easier. It is available from http://sysmod.wordpress.com/2013/03/06/excel-2013-inquire-addin-improved-iq/

The 'Array Formulas' sheet is interesting as a separate list, but I would prefer them to be simply distinguished in the 'All Formulas' sheet by a separate indicator that the formula is an array. Currently they are recognisable by being enclosed in braces {}.

The 'Error Formulas' sheet lists those cells with #Error values. An overflow error, such as a negative or excessive date value, is not reported in this sheet but in 'All Formulas' as a real date (eg 14/04/1791 01:17:02) rather than the ##### error value.

I'm not clear what value there is in listing Logical Formulas, Numeric Formulas, Date/Time Formulas, and Textual Formulas; but somebody probably wants it. My preference is for sheets to be given a map with a specific colour for each data type.

'Numeric Constant Formulas', or 'Textual Constant Formulas' that is, cells with constants embedded ('hardcoded') in the formulas, are a useful list to check for (eg) adjustment values that should be moved to their own cells.

'Nested Ifs' shows formulas with more than one level of IF nesting. This is a useful indicator of a common cause of difficulty in understanding formulas with multiple logic paths. It would be even nicer if the depth of nesting was shown. It would be even better if the complexity measure was applied to All Formulas, rather than just nested Ifs.

'No Cell Refs' is a list of calculations like =1+2+3 but they can also be system functions such as =NOW().

'Blank Cell Refs' is one of these unavoidable reports where most of the information is not useful but buried in there are some possible problem formulas. Excel's error checking has that option turned off by default. Again, I prefer to see these highlighted in context by a mapping tool. Like all auditing tools, all that can be done is to raise the question and leave the judgment to the reviewer.

'Hidden Cells Refs' is potentially of interest, and again a judgement call as to whether the reason for hiding the source cells is for readability (hiding detail not normally needed), analysis (eg grouping or filtering) or obscurity (potential deception).

'Text Cells Refs' can be simply formulas that copy or transform text values, but this report may also pick up the problem where a sum range includes text cells by mistake, or cells accidentally formatted as text.

'External Workbook Refs' is a useful place to collect all the external references. It is clever enough to include external links embedded in defined names. For some reason it also reports some internal cross-sheet references; for example:
='Nov''06'!G5 is reported, but
='Oct 06'!F4 is not reported

'Formatted As Text' can be useful to pick up formulas that are not calculating because they were entered into cells formatted as text. Sometimes people temporarily remove formulas by prefixing them with an apostrophe and then forget to reinstate them.

'Positive Formulas' eg =+B1 may be a sign of old Lotus 1-2-3 habits, where users begin a formula entry with a plus sign rather than an equals sign but are not normally seen as a problem.

'Negative Formulas' eg =-B1 could be an entry made with an initial minus sign or potentially be a keyboard 'fat finger' error because the minus key is beside the equals key on the keyboard. It's not one I've ever seen but there's always a first time.

'Unique Formulas' is probably the most useless listing there is, because all the formulas are regarded as unique in A1 style. I think they simply didn't understand how the term 'unique formula' is normally used in auditing. It is used to describe formulas which are unique in R1C1 style. Other terms for these are 'root formulas' which are copied in blocks, or 'schemas'.

'Duplicate Formulas' shows formulas which are identical on the same sheet in A1 style. This is actually a correct definition of 'duplicate' in this context. It may be useful in order to show several references to the same cell, typically a parameter or title.

'Inconsistent Formulas' are those reported by Excel in its Error Checking rules. It includes those that the user has chosen to ignore and so suppress the green triangle indicator. The following table has the complete list of Excel's rules:

| Excel Error Checking rule | Where reported in Analysis |
| --- | --- |
| Cells containing formulas that result in an error | 'Formulas With Errors' |
| Inconsistent calculated column formula in tables | n/a |
| Cells containing years represented as 2 digits | Not reported |
| Numbers formatted as text or preceded by an apostrophe | 'Numerics As Text' |
| Formulas inconsistent with other formulas in the region | 'Inconsistent Formulas' |
| Formulas which omit cells in a region | n/a |
| Unlocked cells containing formulas | n/a |
| Formulas referring to empty cells | 'Blank Cell Refs' |
| Data entered in a table is invalid | n/a |

Other cell checks that I would like to see are:
- Overflow error
- Formula too long
- Formula with double minus
- Numeric text right aligned
- Range_Lookup missing the fourth parameter
- Formula Hidden
- Formula Unlocked

'Cells With Dependents' lists every cell that is further referenced by other cells. I don't see the value in this. I prefer a colour map that gives a specific colour to each number of dependents, so that if you see a cell that has five dependents in a block where the surrounding cells have four or six, you have something to check.

'Textual Constants' and ' Numeric Constants' list every input value in the workbook. This may have a value for decision models with a few inputs and many calculations, to guide in knowing what cells could be grouped together in an input area.

The 'Comments' sheet lists the sheet name, cell address, formula, value, comment author and text. It may be useful to indicate what people thought useful as meta-data, as information about the spreadsheet formulas and observations on the results. It may reveal previous users and obsolete comments too.

'Validation Criteria' and 'Conditional Formatting' are useful lists as they make obvious structure which is normally under the surface. However, it does not list the actual conditional formatting rules, merely the number of formats applied to each cell.

'Numerics As Text' is also one of Excel's error-checking rules. It is particularly useful when reviewing imported data

'Invisible Cells' are those in hidden rows or columns. It does not report cell contents hidden by formatting codes such as ";;;" or white text, which other auditing tools pick up as bad smells.

'Used Input Cells' are all those which are referred to by formulas. 'Unused Input Cells' are those not referenced by formulas. It does not recognise cells referred to indirectly by OFFSET or INDIRECT.

I can't see the value in a simple list. A colouring map in place would indicate more usefully areas or cells in or beside other used areas, that ought (or ought not) to be referenced.

In this map from XLTest, most cells in the table have 2 dependents (coloured yellow) but two rows have a different colour which means they have only one dependent. This is because the formula to their right omits those cells.

| Number of plays per week: | | | | |
| 1 | 2 | 3 | 4 | Total |
|---|---|---|---|---|
| 311 | 117 | 207 | 301 | 938 |
| 132 | 252 | 272 | 272 | 928 |
| 262 | 282 | 234 | 125 | 883 |
| 172 | 198 | 189 | 234 | 793 |
| 198 | 132 | 189 | 172 | 639 |
| 189 | 156 | 96 | 198 | 639 |
| 181 | 172 | 164 | 76 | 593 |
| 148 | 189 | 76 | 110 | 523 |
| 125 | 117 | 117 | 110 | 469 |
| 103 | 117 | 164 | 76 | 460 |
| 117 | 96 | 64 | 132 | 409 |
| 125 | 70 | 82 | 103 | 300 |
| 82 | 70 | 76 | 64 | 292 |
| 103 | 64 | 46 | 76 | 289 |
| 41 | 36 | 82 | 58 | 217 |
| 64 | 31 | 70 | 46 | 211 |
| 36 | 27 | 70 | 41 | 174 |
| 46 | 27 | 58 | 22 | 126 |
| 31 | 36 | 31 | 27 | 125 |
| 22 | 18 | 27 | 22 | 89 |
| 14 | 27 | 22 | 18 | 81 |
| 11 | 8 | 8 | 11 | 38 |
| 5 | 5 | 8 | 11 | 29 |
| 2 | 2 | 2 | 2 | 8 |
| 1 | 1 | 1 | 1 | 4 |
| 2521 | 2230 | 2355 | 2308 | 9307 |

alty    €465.35

'Merged Cells' may be useful because references to merged areas, or a copy/paste of them, are known to be problematical.

'Occupied Cells' is a list of every value and formula in the workbook. Can anyone think of a use for this? Comparing two versions maybe? Its complement is 'Blank Cells', all the unused cells. What on earth is that for?

'Blank Referenced Cells' may be useful to indicate missing values but again I'd much prefer to see that in context rather than a list.

'Unlocked Cells' indicates that the user has thought of input cells and protection. But the Inquire add-in does not show when sheets are protected. And it does not list separately unprotected cells with formulas, which is one of Excel's error-checking rules. Neither does it list formula cells with the 'Hidden' protection checked. 'Hidden Rows and Columns' is a useful list, although curiously it is sorted by range address regardless of sheet name. It reports rows or columns fully hidden but not very narrow ones which are visually hidden.

'Named Items' only lists the visible names, not the hidden ones. And it does not highlight which names are duplicated with local and global scope, although if they are visible it reports in the Warnings sheet that they exist ('Workbook contains duplicate named ranges.')

'Named Items With Errors' is fair enough, although I'd simply show that as a filter on the full Named Items list.

The final 'Warnings' sheet gives miscellaneous other findings, although there is no documented list of what it is capable of reporting. When I asked that on the Microsoft forums, I was asked "Can you write a justification for this idea? Why document the message list is necessary." They found it hard to believe I actually wanted to know what this add-in could do! As far as I can tell from internal evidence, the possible warnings are:

1. Workbook is setup to change results to same precision as display.
2. Workbook contains formulas with errors.
3. Workbook contains hidden rows or columns.
4. Workbook contains hidden sheets.
5. Workbook contains invisible cells.
6. Workbook contains unlocked cells.
7. Workbook contains duplicate named ranges.
8. Workbook Title has not been set.
9. Workbook Author has not been set.
10. Workbook is setup for R1C1 reference style.
11. Workbook contains sheet names with leading and/or trailing blanks. The sheet names are: . . .

## COMPARISON OF REPRESENTATIONS OF REPEATED FORMULAS

Inquire:

| | A | B | C | D | E |
|---|---|---|---|---|---|
| 1 | **All Formulas (151 total)** | | | | |
| 2 | F:\docs\SCC3\TestWB\PlayTime5.xls | | | | |
| 3 | | | | | |
| 4 | **Sheet Name** | **Cell Address** | **Formula** | **Value** | **Reviewer Comments** |
| 45 | Oct 06 | G11 | =SUM(C11:F11) | 936 | |
| 46 | Oct 06 | G12 | =SUM(C12:F12) | 928 | |
| 47 | Oct 06 | G13 | =SUM(C13:F13) | 883 | |
| 48 | Oct 06 | G14 | =SUM(C14:F14) | 793 | |
| 49 | Oct 06 | G15 | =SUM(C15:F15) | 691 | |
| 50 | Oct 06 | G16 | =SUM(C16:F16) | 639 | |
| 51 | Oct 06 | G17 | =SUM(C17:F17) | 593 | |
| 52 | Oct 06 | G18 | =SUM(C18:F18) | 523 | |
| 53 | Oct 06 | G19 | =SUM(C19:F19) | 469 | |
| 54 | Oct 06 | G20 | =SUM(C20:F20) | 460 | |
| 55 | Oct 06 | G21 | =SUM(C21:F21) | 409 | |
| 56 | Oct 06 | G22 | =SUM(C22:F22)-80 | 300 | |
| 57 | Oct 06 | G23 | =SUM(C23:F23) | 292 | |
| 58 | Oct 06 | G24 | =SUM(C24:F24) | 289 | |
| 59 | Oct 06 | G25 | =SUM(C25:F25) | 217 | |
| 60 | Oct 06 | G26 | =SUM(C26:F26) | 211 | |
| 61 | Oct 06 | G27 | =SUM(C27:F27) | 174 | |
| 62 | Oct 06 | G28 | =126 | 126 | |
| 63 | Oct 06 | G29 | =SUM(C29:F29) | 125 | |
| 64 | Oct 06 | G30 | =SUM(C30:F30) | 89 | |
| 65 | Oct 06 | G31 | =SUM(C31:F31) | 81 | |

Data Connections / Visible Sheets / Hidden Sheets / Very Hidden Sheets / **All Formulas** / Array Form...

This is easy enough for twenty formulas, less so for hundreds or thousands.

Compare that with the Operis Analysis Kit (OAK) Review listing and map:

The map shows in red the beginning of each new block
of a distinct formula.

Another kind of map is illustrated by the XLTest Distinct Formula listing:

| | A | B | C | D | E | F | G | H |
|---|---|---|---|---|---|---|---|---|
| 1 | Distinct | Formula R1C1 in 'Oct 06' | Formula A1 | Count | Areas | Ranges | | |
| 2 | 1 | =SUM(RC[-4]:RC[-1]) | =SUM(C11:F11) | 23 | 3 | ]Oct 06'!G11:G21 | !G23:G27 | !G29:G35 |
| 3 | 2 | =SUM(RC[-4]:RC[-1])-80 | =SUM(C22:F22)-80 | 1 | 1 | ot.xls]Oct 06'!G22 | | |

Which gives the colour key for the XLTest Distinct Formula map:

| | Song Title | Number of plays per week: | | | | |
|---|---|---|---|---|---|---|
| | | 1 | 2 | 3 | 4 | Total |
| 1 | Yellow Nautic | 311 | 117 | 207 | 301 | 936 |
| 2 | Very Fancy | 132 | 252 | 272 | 272 | 928 |
| 3 | Unbelieveable Wanderer | 262 | 262 | 234 | 125 | 883 |
| 4 | Zigbee Follies | 172 | 198 | 189 | 234 | 793 |
| 5 | What Was That About | 198 | 132 | 189 | 172 | 691 |
| 6 | Say You Can't Say What | 189 | 156 | 96 | 198 | 639 |
| 7 | Rich Rio River Rap | 181 | 172 | 164 | 76 | 593 |
| 8 | Questioning, Asking, Wondering | 148 | 189 | 76 | 110 | 523 |
| 9 | The Long and Twisting Way | 125 | 117 | 117 | 110 | 469 |
| 10 | Pay a Packet | 103 | 117 | 164 | 76 | 460 |
| 11 | Old Times Best Forgotten | 117 | 96 | 64 | 132 | 409 |
| 12 | More for the Road | 125 | 70 | 82 | 103 | 300 |
| 13 | Never say maybe | 82 | 70 | 76 | 64 | 292 |
| 14 | Larry the Louche Lizard | 103 | 64 | 46 | 76 | 289 |
| 15 | Kasbah Kismet | 41 | 36 | 82 | 58 | 217 |
| 16 | Jump the Rabbit | 64 | 31 | 70 | 46 | 211 |
| 17 | In Times Gone By | 36 | 27 | 70 | 41 | 174 |
| 18 | Happy Horroreen | 46 | 27 | 58 | 22 | 126 |
| 19 | Go for Gold | 31 | 36 | 31 | 27 | 125 |
| 20 | Fantastic Journeyer | 22 | 18 | 27 | 22 | 89 |
| 21 | Ever in the Future | 14 | 27 | 22 | 18 | 81 |
| 22 | Do what you do be do | 11 | 8 | 8 | 11 | 38 |
| 23 | Can't walk without you | 5 | 5 | 8 | 11 | 29 |
| 24 | Bertie's Blues | 2 | 2 | 2 | 2 | 8 |
| 25 | All about the house | 1 | 1 | 1 | 1 | 4 |
| | Totals | 2521 | 2230 | 2355 | 2308 | 9307 |
| | Royalty | | | | | €465.35 |

**COMPARE THE LISTING OF DEFINED NAMES**

Inquire misses the hidden names (in this case 'Oct 06'!Royalty, see later):

| | A | B | |
|---|---|---|---|
| 1 | *Named Items (4 total)* | | |
| 2 | F:\DOCS\SCC3\TestWB\PlayTime5 unprot.xls | | |
| 3 | | | |
| 4 | **Name** | **Value** | **Reviewer** |
| 5 | Playlist3 | ='Dec-06'!$B$11:$B$35 | |
| 6 | Royalty | =0.071*'F:\DOCS\MyDocs\[Currencies.xls]XE'!$D$7 | |
| 7 | 'Errors'!x | =Errors!#REF! | |
| 8 | 'ECB EuroFXref'!index.en | ='ECB EuroFXref'!$A$1:$I$220 | |
| 9 | | | |

| ◄ ► ►|  | Hidden Rows and Columns | **Named Items** | Named Items With Errors | Warnings |

OAK shows the hidden names and gives useful information on the dimensions of the ranges referred to, to help in consistency checking. Another sheet reports names with overlapping ranges.

| | A | B | C | D | E | F | G | H | I | J | K | L | M |
|---|---|---|---|---|---|---|---|---|---|---|---|---|---|
| 1 | | | There are 11 names defined in Workbook: PlayTime5 unprot.xls | | | | | | | | | | |
| 2 | | | Test carried out: 2013-02-26 15:08:33 | | | | | | | | | | |
| 3 | | | | | | | | | | | | | |
| 4 | Full Name | Short Name | Defined on Sheet | Formula | Refers to Sheet | Address | Area | Top | Bottom | Height | Left | Right | Width |
| 5 | 'ECB EuroFXref'!index.en | index.en | ECB EuroFXref | ='ECB EuroFXref'!$A$1:$I$220 | ECB EuroFXref | $A$1:$I$220 | 1 | 1 | 220 | 220 | 1 | 9 | 9 |
| 6 | Playlist3 | Playlist3 | | ='Dec-06'!$B$11:$B$35 | 06-Dec | $B$11:$B$35 | 1 | 11 | 35 | 25 | 2 | 2 | 1 |
| 7 | Royalty | Royalty | | =0.071*'F:\DOCS\MyDocs\[Currencies.xls]XE'!$D$7 | *** Formula Reference *** | | | | | | | | |
| 8 | 'Oct 06'!Royalty | Royalty | 06-Oct | ='Oct 06'!$B$39 | 06-Oct | $B$39 | 1 | 39 | 39 | 1 | 2 | 2 | 1 |
| 9 | Errors!x | x | Errors | =Errors!#REF! | *** Error Reference *** | | | | | | | | |
| 10 | 'Nov'06'!Z_062E4F23_7EA | Z_062E4F23_7EA0 | Nov06 | ='Nov'06'!$A:$A | Nov06 | $A:$A | 1 | 1 | 65536 | 65536 | 1 | 1 | 1 |
| 11 | 'Dec-06'!Z_062E4F23_7EAI | Z_062E4F23_7EA0 | 06-Dec | ='Dec-06'!$36:$36 | 06-Dec | $36:$36 | 1 | 36 | 36 | 1 | 1 | 256 | 256 |
| 12 | 'Nov'06'!Z_062E4F23_7EA | Z_062E4F23_7EA0 | Nov06 | ='Nov'06'!$1:$1 | Nov06 | $1:$1 | 1 | 1 | 1 | 1 | 1 | 256 | 256 |
| 13 | 'Nov'06'!Z_DBFAFBC0_747 | Z_DBFAFBC0_7477 | Nov06 | ='Nov'06'!$A:$A | Nov06 | $A:$A | 1 | 1 | 65536 | 65536 | 1 | 1 | 1 |
| 14 | 'Dec-06'!Z_DBFAFBC0_74 | Z_DBFAFBC0_7477 | 06-Dec | ='Dec-06'!$36:$36 | 06-Dec | $36:$36 | 1 | 36 | 36 | 1 | 1 | 256 | 256 |
| 15 | 'Nov'06'!Z_DBFAFBC0_747 | Z_DBFAFBC0_7477 | Nov06 | ='Nov'06'!$1:$1 | Nov06 | $1:$1 | 1 | 1 | 1 | 1 | 1 | 256 | 256 |
| 16 | | | | | | | | | | | | | |

| ◄ ► | **Name Database** | Overlapping Ranges | ⊕ |

XLTest shows the visibility and current value:

| | Name | Defined Name | Refers to | Local to sheet | Visible | Type | Value/SUM |
|---|---|---|---|---|---|---|---|
| 55 | | | | | | | |
| 56 | 1 | index.en | ='ECB EuroFXref'!$A$1:$I$220 | 'ECB EuroFXref' | True | Range | 0 |
| 57 | 2 | Playlist3 | ='Dec-06'!$B$11:$B$35 | | True | Range | 0 |
| 58 | 3 | Royalty | ='Oct 06'!$B$39 | 'Oct 06' | False | Range | 0.05 |
| 59 | 4 | Royalty | =0.071*'F:\DOCS\MyDocs\[Currencies.xls]XE'!$D$7 | | True | Error | #REF! |
| 60 | 5 | x | =Errors!#REF! | Errors | True | Error | #REF! |
| 61 | 6 | Z_062E4F23_7E | ='Nov'06'!$A:$A | 'Nov'06' | False | Range | 0 |
| 62 | 7 | Z_062E4F23_7E | ='Dec-06'!$36:$36 | 'Dec-06' | False | Range | 0 |
| 63 | 8 | Z_062E4F23_7E | ='Nov'06'!$1:$1 | 'Nov'06' | False | Range | 0 |
| 64 | 9 | Z_DBFAFBC0_7 | ='Nov'06'!$A:$A | 'Nov'06' | False | Range | 0 |
| 65 | 10 | Z_DBFAFBC0_7 | ='Dec-06'!$36:$36 | 'Dec-06' | False | Range | 0 |
| 66 | 11 | Z_DBFAFBC0_7 | ='Nov'06'!$1:$1 | 'Nov'06' | False | Range | 0 |
| 67 | | | | | | | |

## WORKBOOK RELATIONSHIP

This is a useful picture of the dependencies in systems of linked workbooks.

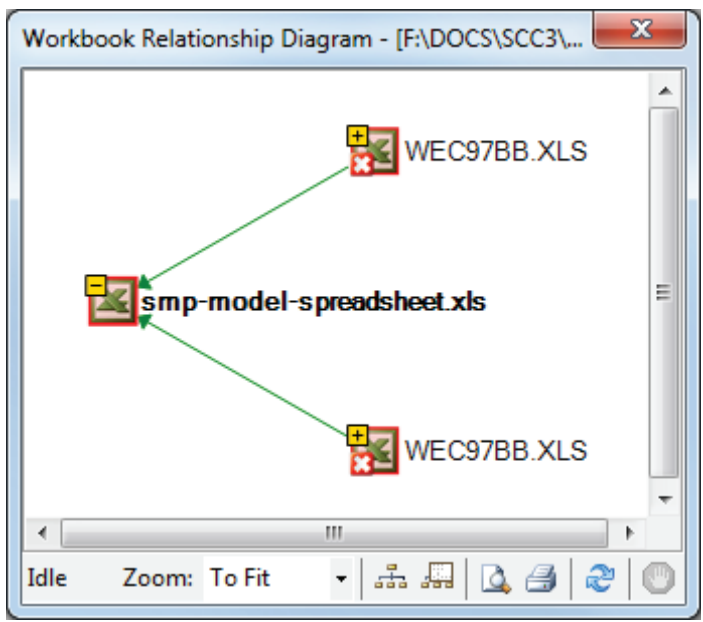

This illustration shows the dead links in the Names. It would be useful if clicking on those linked files gave more information, such as where the links were found.

## WORKSHEET RELATIONSHIP DIAGRAM

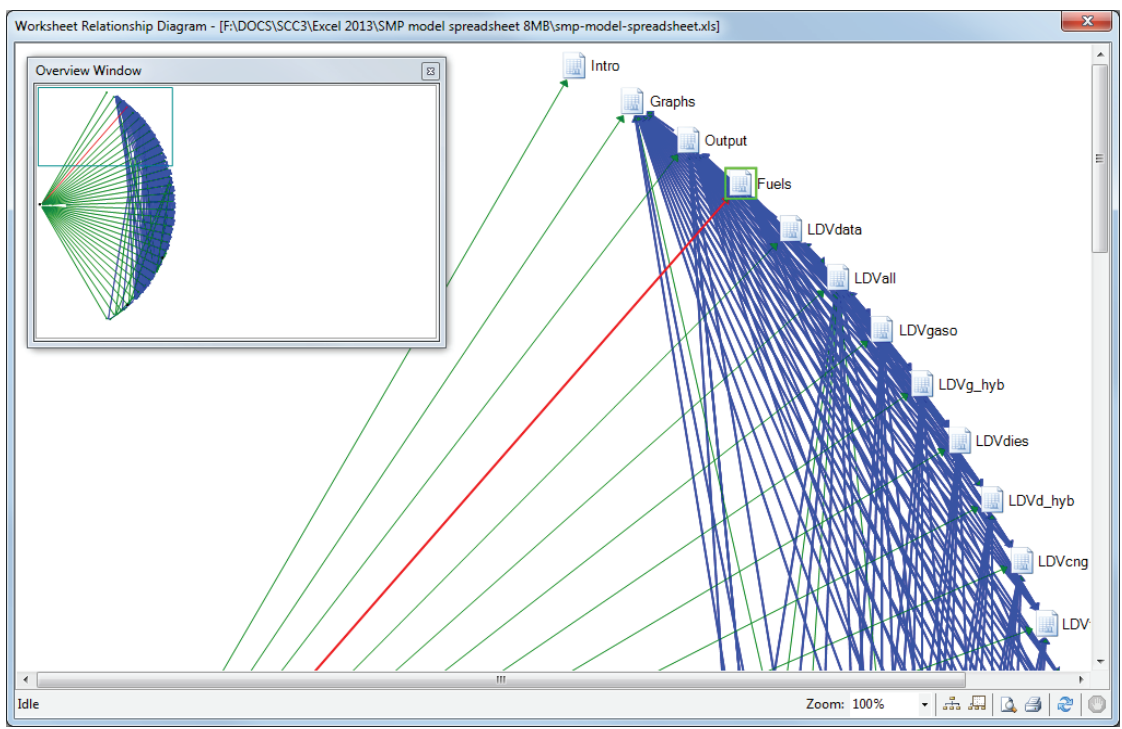

For any moderately complex workbook, this does not provide much enlightenment.
I don't see the value of showing the workbook on the left hand side and all the links going to that, as we already know that they are all in that workbook. It would be clearer to show for example "Intro" on its own unlinked to anything.
Also, I would like the arrow to be annotated with information such as how many links there are for one arrow.

## CELL RELATIONSHIP

This shows the links between individual cells. Because it does not have any concept of ranges of similar formulas, the diagrams rapidly sprawl as the number of copied formulas increases. Here is an image from a simple tax calculation spreadsheet:

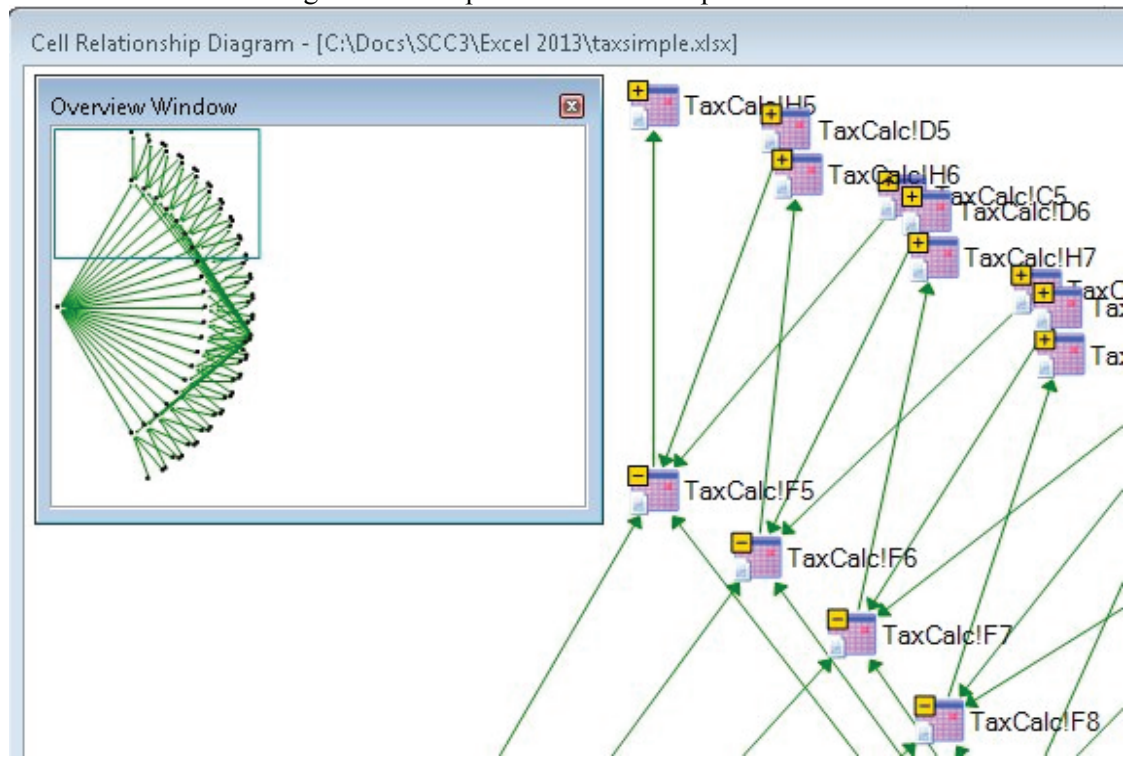

It is therefore probably best suited to simple spreadsheets.

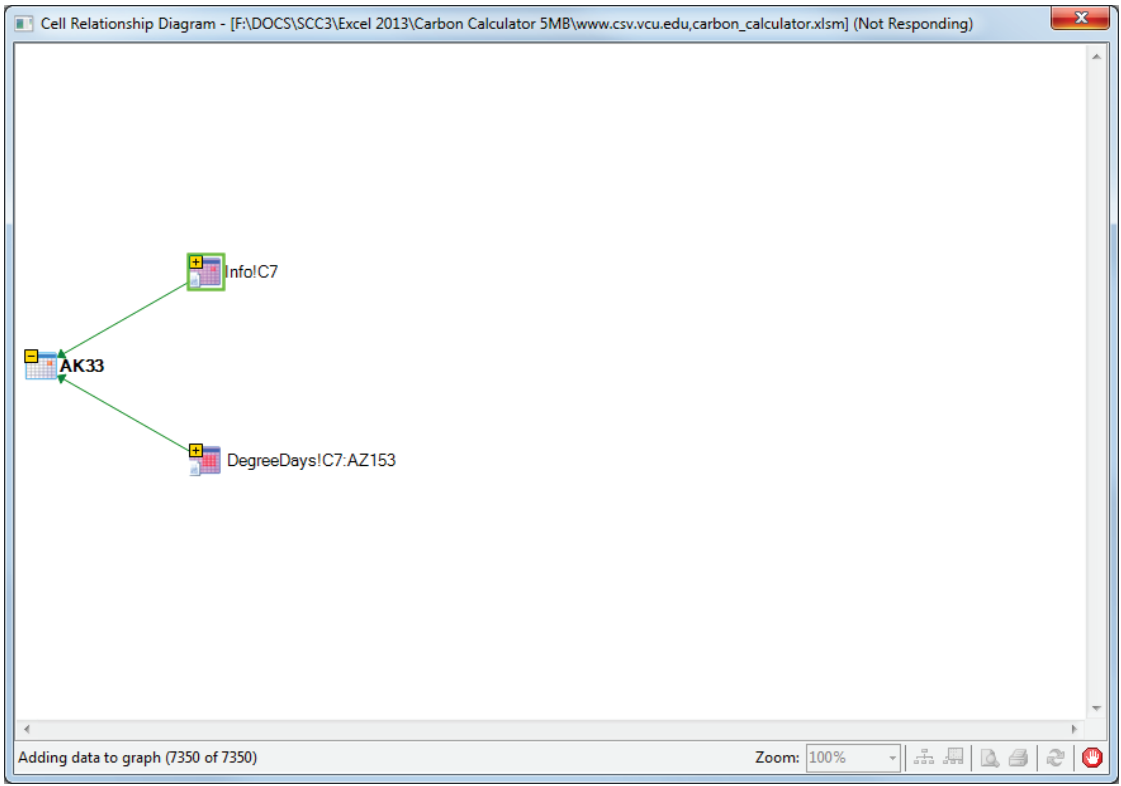

It will try to expand the tree for all the precedent cells in a LOOKUP formula, which could be thousands. This can cause Excel to hang (the window goes frosty and the title bar says "(Not Responding)" for several minutes. Task Manager can show you how it using memory and CPU time:

| Image Name | User Name | CPU | Memory (... | Description |
|------------|-----------|-----|-------------|-------------|
| EXCEL.EXE | Patrick | 25 | 390,780 K | Microsoft Excel |

Eventually it will give you a graph with thousands of nodes:

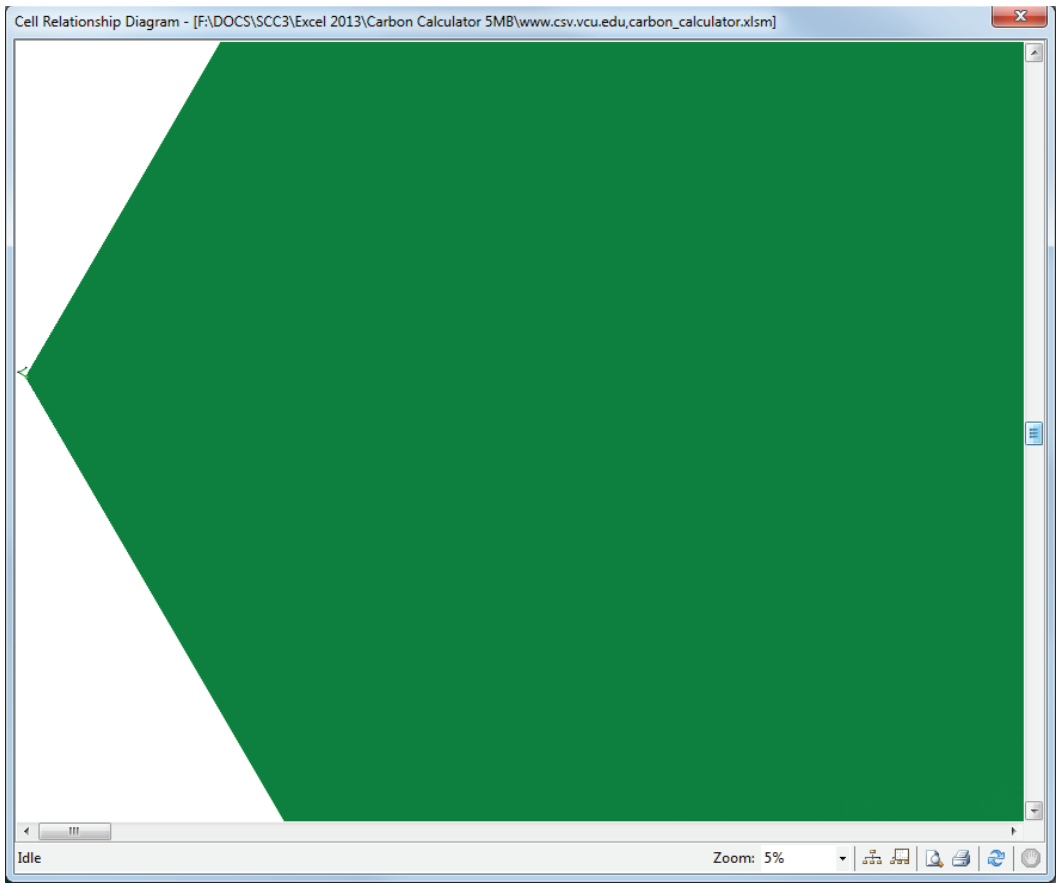

## COMPARE FILES

The Compare Files command lets you see the differences, cell by cell, between two workbooks. Results are colour-coded by the kind of content, such as entered values, formulas, named ranges, and formats. A window shows VBA code changes line by line. Differences between cells are shown in a grid layout, like this:

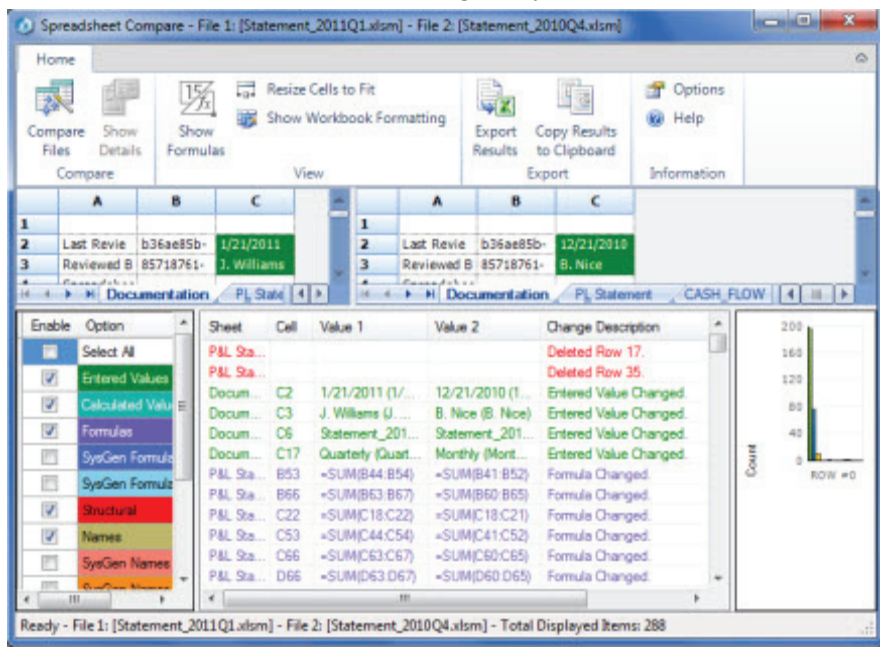

The "Show Details" button applies to VBA differences and shows a WinDiff output of the code. Clicking "Help" just gets you
"We're sorry, but there is no help available for Spreadsheet Compare in this market."

There is a rather obscure term 'SysGen …' in the options that is not explained anywhere in the Help. It only makes a difference to the report after you tick the option 'Include system generated changes in result' in the Options dialog which appears when you start a Compare; so you have to do at least one comparison, even if you do not use the output, to get at the following dialog: (As you can unselect the four SysGen checkboxes in the Options, I don't know why they don't always process it by default as you can always tick or untick the display later.)

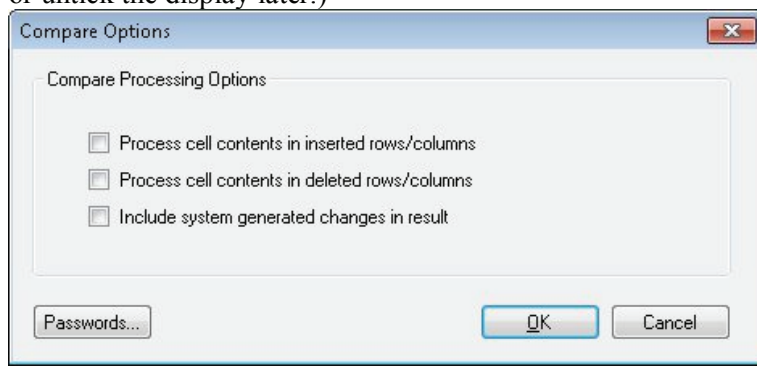

By 'system generated changes' they mean formulas that have been automatically changed by Excel as a result of an insertion or deletion. For example, say you have a formula =SUM(E6:E9) and you delete row 9 so the formula now shows =SUM(E6:E8).

With 'SysGen Formulas' unticked, no difference is reported:

When ticked, the difference is highlighted:

'SysGen Names' refers to Excel's internal names such as Print_Area.

**BUG IN THE COMPARE OPTIONS LIST**

If you click the Option heading in the Option checkboxes at the bottom left, the table is sorted alphabetically by Option. But the first checkbox is still "Select All"; that is, the action of the checkboxes does not change from the unsorted list.

The 'Entered Values' comparison is case sensitive, and there does not appear to be an option to make it treat upper and lower case as being the same.

The following page shows a complete picture of the Compare windows. If exported to Excel, the comparison is shown as a table like this:

| Sheet | Range | Old Value | New Value | Description |
|---|---|---|---|---|
| Staffin g Plan | | | | Added Column K. |
| Staffin g Plan | | | | Deleted Row 9. |
| Staffin g Plan | | | | Deleted Row 11. |
| Staffin g Plan | I19 | =ROUND($C$21/I 19,-3)/4/1000 | =ROUND($C$19/I 17/1000,0)/4 | Formula Changed. |
| Staffin g Plan | J19 | =ROUND($C$21/J 19,-3)/4/1000 | =ROUND($C$19/J 17/1000,0)/4 | Formula Changed. |
| Staffin g Plan | C19 | 350,000 (350000) | 400,000 (400000) | Entered Value Changed. |
| Staffin g Plan | E19 | 10 (9.75) | 11 (11) | Calculated Value Changed. |
| Staffin g Plan | F19 | 7 (6.75) | 8 (7.75) | Calculated Value Changed. |
| Staffin g Plan | G19 | 5 (5.25) | 6 (6) | Calculated Value Changed. |
| Staffin g Plan | H19 | 4 (4.25) | 5 (4.75) | Calculated Value Changed. |
| Staffin g Plan | I19 | 4 (3.5) | 4 (4) | Calculated Value Changed. |
| Staffin g Plan | J19 | 3 (3) | 4 (3.5) | Calculated Value Changed. |
| Staffin g Plan | _xlnm.Pri nt_Area | ='Staffing Plan'!$C$4:$J$21 | ='Staffing Plan'!$C$4:$J$19 | Named Item: _xlnm.Print_Area Definition Changed By System. |

For display purposes OAK can insert rows or columns to line up the sheets to match:

| Range | Cell contents on 1st Sheet | Cell contents on 2nd Sheet |
|---|---|---|
| From Row = 9 | | 1 rows inserted |
| From Row = 11 | | 1 rows inserted |
| From Column = K | 1 columns inserted | |
| K1 | | Staffing |
| K2 | | Q3 |
| K3 | | Year 2 |
| K6 | | 1 |
| K7 | | 20 |
| K8 | | 4 |
| B9 | Other | |
| C9 | Input | |
| E9 | 0 | |
| F9 | 0 | |
| G9 | 0 | |
| H9 | 0 | |
| I9 | 0 | |
| J9 | 0 | |
| L9 | 80000 | |
| E10 | =SUM(E6:E9) | =SUM(E6:E8) |
| F10 | =SUM(F6:F9) | =SUM(F6:F8) |
| G10 | =SUM(G6:G9) | =SUM(G6:G8) |
| H10 | =SUM(H6:H9) | =SUM(H6:H8) |
| I10 | =SUM(I6:I9) | =SUM(I6:I8) |
| J10 | =SUM(J6:J9) | =SUM(J6:J8) |
| K10 | | =SUM(K6:K8) |
| K13 | | 1 |
| K14 | | 3 |
| K15 | | 2 |
| K16 | | 2 |
| K17 | | =SUM(K13:K16) |
| K19 | | =K10+K17 |
| C21 | 350000 | 400000 |
| I21 | =ROUND($C$21/I19,-3)/4/1000 | =ROUND($C$21/I19/1000,0)/4 |
| J21 | =ROUND($C$21/J19,-3)/4/1000 | =ROUND($C$21/J19/1000,0)/4 |
| K21 | | =ROUND($C$21/K19/1000,0)/4 |

**CLEAN EXCESS CELL FORMATTING**

This says it clears excessive formatting but does not tell you beforehand what will be cleared, nor afterwards what was cleared.

**Interactive analysis**

This was removed from the final release of the Inquire add-in for Excel 2013, because of stability problems.

By comparison, OAK has a menu to report various items of interest:

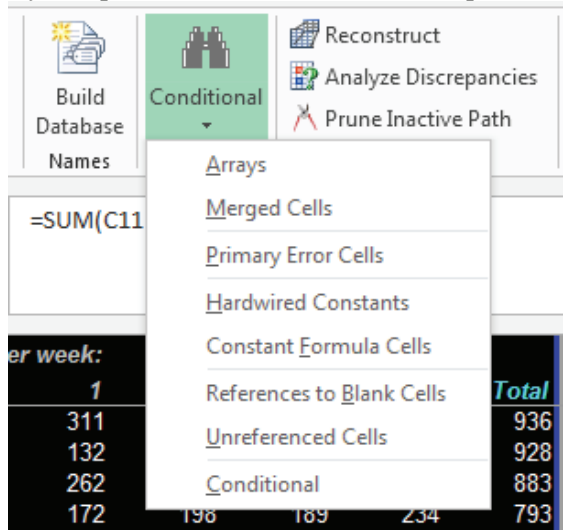

A strong feature of OAK is its facility to understand the reconstruct the components of a formula and to progressively simplify it, for example by pruning inactive paths.

Again to compare with another tool, XLTest produces an analysis either interactively or as a batch process:

| | Sheet | Type | Sheet Tab Name | Visibility | Contents | Sum Total | Rows | Columns |
|---|---|---|---|---|---|---|---|---|
| 2 | 1 | Worksheet | $TOC(1) | Visible | | | | |
| 3 | 2 | Worksheet | $Nuf Front (1) | Visible | Number Format | 46 | 6 | 9 |
| 4 | 3 | Worksheet | $Sty Front (1) | Visible | Styles | 15 | 2 | 7 |
| 5 | 4 | Worksheet | $Nuf Dec-06(1) | Visible | Number Format | 19 | 4 | 7 |
| 7 | 6 | Worksheet | $DpC Consolidated(1) | Visible | Dependents Count in 'Consolidated' | 0 | 1 | 1 |
| 15 | 14 | Worksheet | $PrL Front (1) | Visible | Precedents Location in 'Front ' | 0 | 1 | 1 |
| 24 | 23 | Worksheet | $Use Front (1) | Visible | Data Type and Usage in 'Front ' | 0 | 1 | 1 |
| 33 | 32 | Worksheet | $Typ Front(1) | Visible | Data Types in 'Front ' | 0 | 1 | 1 |
| 42 | 41 | Worksheet | $Val Front(1) | Visible | Data Validation | 3 | 2 | 6 |
| 43 | 42 | Worksheet | $For Front (1) | Visible | Distinct Formulas | 152 | 16 | 6 |
| 52 | 51 | Worksheet | $Cof Nov'06(1) | Visible | Conditional Formats | 9 | 2 | 6 |
| 53 | 52 | Worksheet | $Nuf Nov'06(1) | Visible | Number Format | 9 | 3 | 7 |
| 56 | 55 | Worksheet | $Sty Front(1) | Visible | Styles | 9 | 3 | 7 |
| 59 | 58 | Worksheet | $Inf _ Front (1) | Visible | Inf_ Inspection of 'Front ' | 15386 | 63 | 6 |
| 70 | 69 | Worksheet | $VBA(1) | Visible | Reference | 2610 | 27 | 7 |
| 71 | 70 | Worksheet | $WSXref(1) | Visible | First reference by each sheet named across the page | 11 | 22 | 14 |
| 72 | 71 | Worksheet | $DOC(1) | Visible | 26/02/2013 | #REF! | 168 | 31 |
| 73 | 72 | Worksheet | $APP(1) | Visible | Excel Application Settings | 42201.49313 | 62 | 7 |
| 74 | 73 | Worksheet | XLTEST_LOG | Visible | Date, time | 11458786455 | 103 | 14 |

XLTEST provides a detailed worksheet analysis:

| Inf_<br>Inspectio<br>n of<br>'Errors' | Worksheet | | Errors |
|---|---|---|---|

No Named ranges created for checks

| No. | Error Value | Count | Areas | Ranges | …. |
|---|---|---|---|---|---|
| 2007 | #DIV/0! | 1 | 1 | | B22 |
| 2042 | #N/A | 1 | 1 | | B16 |
| 2029 | #NAME? | 2 | 2 | B13 | D23 |
| 2000 | #NULL! | 1 | 1 | | B25 |
| 2036 | #NUM! | 2 | 2 | C5 | B23 |
| 2023 | #REF! | 3 | 3 | B17 | B21 |
| 2015 | #VALUE! | 2 | 2 | B11 | B15 |

| No. | Error Check | Count | Areas | Ranges | |
|---|---|---|---|---|---|
| 1 | Error value | 12 | 8 | C5 | B11 |
| 2 | Text two digit year | | | | |
| 3 | Number stored as text | 1 | 1 | | F20 |
| 4 | Inconsistent formula | | | | |
| 5 | Omits cells in region | | | | |
| 6 | Unlocked formula cell | | | | |
| 7 | Refers to empty cell | 4 | 4 | B16 | B22 |
| 8 | List validation error | | | | |
| 9 | Inconsistent list | | | | |
| 10 | Fails data validation | | | | |
| 11 | Overflow error | 2 | 2 | B5 | E5 |
| 12 | Number in formula | 6 | 4 | E5 | B11 |
| 13 | Format hides value | | | | |
| 14 | Format Font Fill colour | | | | |
| 15 | Conditional Format Font | | | | |
| 16 | Formula too long | | | | |
| 17 | Formula starts with minus | 1 | 1 | | B5 |
| 18 | Formula with double minus | | | | |
| 19 | Numeric text right aligned | 1 | 1 | | F20 |
| 20 | Range_Lookup check params | 3 | 1 | | B15:B17 |
| 21 | FormulaLabel property | | | | |

| Other | Checks | Count | Address | |
|---|---|---|---|---|
| 35 | Cells Checked | 210 | A1:F30 | |
| 36 | Not empty | 64 | | |
| 37 | Formulas | 32 | | |
| 38 | Distinct Formulas | 27 | | |
| 39 | Formula Hidden | 0 | | |
| 40 | Formula Unlocked | 0 | | |
| 41 | Longest Formula | 102 | E20 | ='F:\DOCS\SCC3\Ex1<br>Demo\[EX1DEMO.X<br>LS]Budget08'!N70+'F |

| No. | | Count | Reference | | |
|---|---|---|---|---|---|
| | | | | | ...\DOCS\SCC3\Ex1Demo\[s3errors.xls]InsRow'!$E$5+B20 |
| 42 | Array Formulas | 1 | 1 | | B27 |
| 43 | Object formulas | 0 | | | |
| 44 | Merged Areas | 6 | 1 | | A1:F1 |
| 45 | Number Formats used | 5 | | | |
| 46 | Custom Styles used | 1 | | | |

| 47 | Circular Reference | $D$23 | Reference | Formula | In Names |
|---|---|---|---|---|---|
| | | | 1 | [PlayTime5 unprot.xls] Errors'!$D$23 | =SUM(C19:D23) |

| No. | Function | Total | Distinct |
|---|---|---|---|
| 1 | NOW | 1 | 1 |
| 2 | DATE | 1 | 1 |
| 3 | IF | 12 | 4 |
| 4 | ISERROR | 8 | 3 |
| 5 | ISERR | 5 | 1 |
| 6 | Total | 1 | 1 |
| 7 | VLOOKUP | 4 | 4 |
| 8 | SUM | 4 | 4 |
| 9 | LOG | 1 | 1 |
| 10 | INDIRECT | 2 | 2 |

| No. | Number Formats used | Count | First use |
|---|---|---|---|
| 1 | General | 203 | A1 |
| 2 | m/d/yyyy h:mm | 1 | B5 |
| 3 | m/d/yyyy | 3 | C5 |
| 4 | 0.00 | 2 | B7 |
| 5 | @ | 1 | D13 |

| No. | Custom Styles used | Count | First use |
|---|---|---|---|
| 1 | Normal 2 | 207 | A1 |

| Comments | Cell note text | Cell Value | Cell Formula |
|---|---|---|---|
| B27 | POB: This array function was entered using Ctrl+Shift+Enter | 9 | =SUM(IF(ISERROR(B1:B25),1,0)) |

**CONCLUSION**

The Inquire add-in has the great advantage of being included with Excel, rather than a separate purchase. It provides some minimum standard of documentation, although the simplistic formula listing is not useful for large workbooks. Commercial auditing software offers much more and for a modest cost, typically less than 400 US dollars. Some, like OAK, go beyond review facilities and offer development assistance as well.

Patrick O'Beirne
Systems Modelling Ltd, Ireland
Landline +353 5394 22294      mobile:+353 86 835 2233
Email: pob@sysmod.com
Systems Modelling website: http://www.sysmod.com
Blog: http://sysmod.wordpress.com
LinkedIn profile: http://ie.linkedin.com/in/patrickobeirne

**REFERENCES**

[1] http://office.microsoft.com/en-us/excel-help/what-you-can-do-with-spreadsheet-inquire-HA102835926.aspx
[2] http://www.spreadsheetdetective.com/
[3] http://www.auditware.co.uk/
[4] http://arxiv.org/abs/0805.4236
[5] http://arxiv.org/abs/0712.2591
[6] http://www.sysmod.com/xltest
[7] http://www.operis.com
[8] http://www.abb.com/consulting
[9] http://www.liquidity.com
[10] http://www.clusterseven.com
[11] http://www.sarbox-solutions.com
[12] http://arxiv.org/abs/0805.4211
[13] http://arxiv.org/abs/1111.5002
[14] http://arxiv.org/abs/0808.2045
[15] http://arxiv.org/abs/0805.1741